\documentstyle[12pt,epsfig]{article}

\columnwidth\textwidth

\def\slash#1{\rlap{$#1$}/ } % slashes a narrow character
\def\bigslash#1{\rlap{$#1$}\thinspace / \thinspace }
\def\comment#1{}

\begin{document}
\pagenumbering{arabic}
%-----------------------------
\everymath={\displaystyle}
\vspace*{-2mm}
\thispagestyle{empty}
%\centerline{\bf --- PRELIMINARY VERSION AS OF \today \ ---}
\noindent
\hfill HUTP-96/A053\\
\mbox{}
\noindent \hfill hep-ph/9701243\\ 
\noindent \mbox{}
\hfill January 24, 1997 \\
\begin{center}
  \begin{Large}
  \begin{bf} {\Large \sc 
Reparameterization Invariance Revisited}
   \\
  \end{bf}
  \end{Large}
  \vspace{0.8cm}
   Markus Finkemeier, 
   Howard Georgi,
   Matt McIrvin\\[2mm]
   {\em Lyman Laboratory of Physics\\
        Harvard University\\
        Cambridge, MA 02138, USA}\\[5mm]
{\bf Abstract} % =====================================================
\end{center}
\begin{quotation}
\noindent
Reparameterization invariance, a symmetry of heavy quark effective
theory, appears in different forms in the literature. The most commonly
cited forms of the reparameterization transformation are shown to induce
the same constraints on operators that do not vanish under the equation
of motion to order $1/m^2$, and to be related by a redefinition of the
heavy quark field. 
We give a new, very straightforward proof that that the reparameterization
invariance constraints apply to all orders in $\alpha_s$ under matching to 
full QCD and renormalization-group running, at least up to and including 
$O(1/m^2)$.
\end{quotation}
%                 ====================================================
%
%\vspace*{1cm}
%
%\centerline{PACS numbers: }
%
%\newpage

%======================================================================
\section{Introduction}
%======================================================================
Heavy particle effective theories are useful in a variety of situations
\cite{hpet1,hpet2,hpet3,hpet4,hpet5,hpet6,hpet7,scalar}. 
In these effective theories, 
$S$ matrix elements are expanded around the limit $1/m \to 0$,
in which limit the heavy particle becomes nearly static and the velocity $v$
of the heavy particle becomes a conserved quantum number. The momentum $p$ of
the heavy particle is decomposed as
\begin{equation}
   p = m v + k
\end{equation}
where $m$ is the mass of the heavy particle, and $v$ is a four velocity 
($v^2=1$), which must be choosen such that the residual momentum $k$ is 
small compared to $m$. 
Clearly, the decomposition $p=mv+k$ is not unique (see \cite{Dug92}, for 
example). We can as well write
$p = m v' + k'$ where $k' = k + (v-v')/m$, as long as $v'^2 = 1$. 

This leads to the requirement of reparameterization invariance for the 
effective Lagrangian \cite{Luk92,Che93,Kil94}. 
In the case of a scalar field $\phi(x)$ 
\cite{scalar,Luk92}, 
the issue is rather 
simple.  Let us consider an infinitesimal reparameterization
\begin{equation} \label{eq:infini}
    v \to v' = v + \delta v \qquad \mbox{where} \qquad v \cdot \delta v = 0
\end{equation}
The effective Lagrangian ${\cal L}_v$ is written in terms of $\phi_v(x)$,
defined by
\begin{equation}
   \phi_v(x) = \sqrt{2m} \exp(i m v\cdot x) \phi(x)
\end{equation}
and the reparameterization (\ref{eq:infini}) leads to
\begin{equation} \label{eq:scalar}
   \phi_v \to \phi_v' = \exp(i m \delta v \cdot x) \phi_v
   = [1 + i m \delta v \cdot x] \phi_v
\end{equation}
Due to the appearance of $m$ in the transformation law (\ref{eq:scalar}), the
requirement of invariance of the effective Lagrangian under reparameterization
leads to relations between couplings of different order in $1/m$.

In the case of spin $1/2$, the situation is more complicated, because
the reparameterization transformation of the field $\Psi_{+v}(x)$ must
involve a rotation in Dirac space, in order to ensure that the projection 
identity $\slash v \Psi_{+v}(x) = \Psi_{+v}(x)$ is transformed into 
$\slash v' \Psi_{+v}(x) = \Psi_{+v'}(x)$. 

Indeed, there is controversy in the literature on the correct form of
the reparameterization transformation for heavy quark effective theory.
In their paper on the issue, Luke and Manohar \cite{Luk92} propose
a certain form for this transformation $\Psi_{+v} \to \Psi_{+v'}$. 
This transformation law has been criticized by Yu-Qi Chen \cite{Che93} as
being incorrect at $O(1/m^2)$. 

Chen proposed a different transformation law, and in fact it is straightforward
to calculate that the effective Lagrangian obtained from tree level matching to
full QCD is invariant under Chen's transformation, but not under the one
proposed
by Luke and Manohar. This alone, however, does not imply that Luke and Manohar's
transformation law is incorrect, because the form of the effective Lagrangian is
not unique. Field redefinitions of the heavy quark field can change the
Lagrangian without changing the physical predictions.

The purpose of this paper is to shed some light on this question. In fact, we
have not been able to follow the arguments in either \cite{Luk92} nor in
\cite{Che93} regarding the derivation of the reparameterization transformation
step by step. To which extent this is due to our own inabilities, and to which
extent the arguments are actually inconclusive or wrong, is not completely clear
to us at each point, either. Therefore we decided to investigate the issue
on our
own along somewhat different lines.

Our main results are as follows. The difference between Chen's
transformation and
Luke and Manohar's, at least to order $1/m^2$, has to do with the presence
of the
``class II operators'' that vanish under the leading-order equation of motion.
They are therefore members of a family of reparameterization transformations
which, interpreted as symmetries, impose the same constraints on the
coefficients
of the ``class I operators'' that do not vanish under the leading-order 
equation
of motion. We have found a new, very straightforward proof that these
constraints
in fact hold to order $1/m^2$ in the heavy mass expansion, not merely at level,
but to arbitrary order $\alpha_s^n$ in QCD perturbation theory. We also prove
that the constraint imposed by Chen's transformation on class II operators 
holds
to order $1/m^2$ but all orders in $\alpha_s$, 
if one uses the field definitions
obtained via our straightforward type of matching. 

One might suspect that analogous relations will hold at higher order $1/m^3$, 
but we have not proven that. Furthermore, it is not really clear to us that
reparameterization invariance constraints will hold unchanged for
non-perturbative effects. We would like to encourage further study of this
issue.

The present paper is organized as follows.

In Sec.~2, we review the structure of the heavy quark effective theory
Lagrangian.
In Sec.~3, we show how Chen's transformation law can easily be derived on a
classical level. We find that the tree level matching Lagrangian
is invariant under this transformation law.
Then,  in Sec.~4, we compare this with Luke and Manohar's transformation. 
We show that the two transformation laws differ by a redefinition
of the fields.

In Sec.~5, we discuss reparameterization invariance constraints on
the couplings of the effective Lagrangian, and a subtlety in applying
the statements in \cite{Luk92} to the relations between the coupling
coefficients at order $1/m^2$. There has been some confusion over the
implications of Luke and Manohar's version of reparameterization
invariance. We show that Luke and Manohar's transformation
actually yields the same class I constraints as Chen's.

In Sec.~6, we discuss loops and matching corrections. The Wilson coefficents
$C_i(\mu)$
which multiply the various operators in the effective Lagrangian can be
obtained in a two-step process. In the first step, ``matching'', one can
use a renormalization scale $\mu=m$. The $C_i(m)$ are then determined by 
requiring Green's functions in the full and the effective theory to be equal.
In the second step, ``running'', the renormalization group equations are 
used to evolve down from $m$ to scales $\mu \ll m$. We discuss
why the invariance under Chen's transformation is preserved to all orders
in $\alpha_s$ and up to (including) order $1/m^2$ in these two steps.
In Sec.~7 we draw our conclusions.

%======================================================================
\section{Operators in the heavy quark Lagrangian}
%======================================================================

The general form of the heavy quark effective
Lagrangian is given by \cite{lee,Mci96}
\begin{eqnarray} \label{eq:eff}
        {\cal L}^{\rm eff} & = & \overline{\Psi}_{+v} i{D \cdot v} \Psi_{+v} + 
        C_{kin} O_{kin} + C_{mag} O_{mag}
       + C_1 O_1 + C_2 O_2 
\nonumber \\ & & {} + \mbox{(class II terms)} + O(1/m^3)
\end{eqnarray}
where
\begin{eqnarray}
        O_{kin} &=& -{1 \over 2m} 
   \overline{\Psi}_{+v} D^2 \Psi_{+v} \nonumber \\
        O_{mag} &=& {g \over 4m} 
   \overline{\Psi}_{+v} \sigma^{\mu \nu} G_{\mu \nu} \Psi_{+v}
                                \nonumber \\
        O_1     &=& {g \over 8m^2} 
   \overline{\Psi}_{+v} v^{\mu} [D^{\nu}, G_{\mu \nu}]
                \Psi_{+v} \nonumber \\
        O_2     &=& {ig \over 8m^2} 
   \overline{\Psi}_{+v} \sigma^{\alpha \mu} v^{\nu}
                \lbrace D_{\alpha},G_{\mu \nu}\rbrace \Psi_{+v}
    \label{eq:operators}
\end{eqnarray}
We define $D_\mu = \partial_\mu - i g A_\mu^a T^a$, and
$G_{\mu \nu} = {i \over g} [D_\mu, D_\nu]$.

We have chosen to define the operators $O_i$ such that tree level
matching to full QCD yields $C_1 = C_2 = C_{kin} = C_{mag} = 1$.
There is more freedom in defining the class I operator basis, which
comes from the ability to add or remove class II terms from these
operators. The definitions above make quantum corrections and
field redefinitions easy to deal with.

Class II operators have the general form
\begin{equation}
   O_i = \overline{\Psi}_{+v} (iD\cdot v A + \overline{A} i D \cdot v)
    \Psi_{+v}   
\end{equation}
and so vanish when applying the classical equations of motion. They
can be removed from the effective Lagrangian by a field redefinition
which does not change the coefficients of the class I operators 
$C_{kin}$, $C_{mag}$, $C_1$ and $C_2$.
A convenient basis of operators to work with is
\begin{eqnarray}
	O_{D \cdot v} &=& -{1 \over 2m} {\overline{\Psi}_{+v}} (i D \cdot v)^2
{\Psi_{+v}} \nonumber \\
	O_a &=& {1 \over 4m^2} {\overline{\Psi}_{+v}} \lbrace i D \cdot v, (iD)^2
\rbrace {\Psi_{+v}} \nonumber \\
	O_b &=& {1 \over 4m^2} {\overline{\Psi}_{+v}} (i D\cdot v)^3 {\Psi_{+v}}
\nonumber \\
	O_c &=& {-g \over 8m^2} {\overline{\Psi}_{+v}} \lbrace i D \cdot v, G^{\mu \nu}
		\sigma_{\mu \nu} \rbrace {\Psi_{+v}}
\end{eqnarray}
Upon tree-level matching, $C_{D \cdot v} = -1$,$C_a = -1/2$, $C_b = 1$, and
$C_c = 1$.
(This basis was chosen to make it easy to translate results about
renormalization-group running from \cite{Bal96}.)

We have used a basis of Hermitian operators. Its most important feature is
that the class I and class II parts of the Lagrangian are separately
Hermitian. This makes it possible to remove class II operators with
field redefinitions, as in \cite{Mci96} and below.

It is purely for convenience that we define the 
{\it individual} operators to be
Hermitian, but if one uses a non-Hermitian operator 
basis and wishes to leave in
the class II part, it is important to include enough operators to be able to
reproduce all of these Hermitian operators by taking linear combinations. 
(See our comments on \cite{CKO95} in Sec. 5.1 below.)

%==========================================================================
\section{Derivation of Chen's transformation law at the classical level}
%==========================================================================

Let's start from the heavy quark effective theory using $v$. One defines
\begin{equation}
   \Psi_{\pm v} = e^{i m v\cdot x} \frac{1\pm \slash v}{2} \Psi(x)
\end{equation}
where $\Psi(x)$ is the quark field that appears in the QCD Lagrangian. 
This implies
\begin{equation}
   \Psi(x)  = e^{-imv\cdot x} [\Psi_{+v}(x) + \Psi_{-v}(x)]
\end{equation}
The tree level matching Lagrangian is obtained by integration
out the  heavy field $\Psi_{-v}$ using the classical equations of motion
\begin{equation}
   \Psi_{-v} = \frac{1}{2 m + i v\cdot D} i ({\bigslash D} - v \cdot D) 
   \Psi_{+v}
\end{equation}
Now consider a effective theory using $v'$ with
\begin{equation} 
    v \to v' = v + \delta v \qquad \mbox{where} \qquad v \cdot \delta v = 0
\end{equation}
We can express $\Psi_{+v'}$ through $\Psi_{+v}$ by using the classical
equations of motion. We have
\begin{eqnarray}
   \Psi_{+v'} & = & e^{i m v' \cdot x} P_{+v'} \Psi
\nonumber \\
\nonumber \\
   & = & e^{i m (v + \delta v)\cdot x}
    \frac{1 + \slash v + \delta \slash v}{2}
   e^{-imv\cdot x} \Big[1 + \frac{1}{2 m + i v\cdot D} 
    i ({\bigslash D} - v \cdot D) \Big] \Psi_{+v}
\nonumber \\
\nonumber \\
   & = & \left[ 1 + i m \delta v \cdot x + 
   \frac{\delta \slash v}{2} + \frac{\delta \slash v}{2}
   \frac{1}{2 m + i v \cdot D} i ({\bigslash D} - v \cdot D) \right] \Psi_{+v}
\label{eq:chen}
\end{eqnarray}
which is Chen's transformation law \cite{Che93} (see also \cite{Kil94}).

In the above derivation of the transformation law,
we have used the classical equations of motion for $\Psi_{-v}$.
It is therefore not obvious whether matching corrections to the effective
Lagrangian will be invariant under the transformation law.

Furthermore, something funny has happened. 
In the effective theory based on $v$, the two 
``heavy components'' $\Psi_{-v}$ are integrated
out, and the two ``light components'' $\Psi_{+v}$ are left as degrees of 
freedom. In the effective theory based on $v'$, slightly different
degrees of freedom, namely $\Psi_{-v'}$ are integrated out. 
One might think that it should not be possible
to recover $\Psi_{+v'}$  from $\Psi_{+v}$ (Note that in the case of a
heavy particle effective theory for a scalar field, these problems do not
appear because in that case there are no degrees of freedom which are
integrated out.).

At the tree level, however, everything is certainly correct.
We have checked explicitly that the tree level matching Lagrangian 
\begin{equation}
   {\cal L }^{\rm tree} = \overline {\Psi}_{+v} 
   \left[
   i v D + i {\bigslash D}_\perp \frac{1}{2 m + i v D} i {\bigslash D}_\perp
   \right]
   \Psi_{+v}
\end{equation}
(expansion in $1/m$ is implied)
is invariant under the transformation law (\ref{eq:chen}).
The calculation is somewhat lengthy, but straightforward. It is given 
here in Appendix A. 

%========================================================================
\section{Comparing with Luke and Manohar's transformation}
%========================================================================

\subsection{The difference between the transformations}
Luke and Manohar propose the following transformation law for the spinor
$\Psi_{+v}$ under reparameterization transformations \cite{Luk92}:
\begin{equation}
   \Psi_{+v}(x) \to \Psi_{+v'}^{\rm LM}(x) 
   = e^{i m \delta v \cdot x} \Lambda(v',\hat{u}) \Lambda(v,\hat{u})^{-1}
    \Psi_{+v}(x)
\end{equation}
where
\begin{equation}
   \hat{u}^\mu = \frac{\displaystyle v^\mu + \frac{i D^\mu}{m}}
   {\sqrt{\displaystyle 1 + \frac{2 i v\cdot D}{m} - \frac{D^2}{m^2}}}
\end{equation}
and
\begin{equation}
   \Lambda(w,v) = \frac{1 + \slash w \slash v}{\sqrt{2(1 + v\cdot w)}}
\end{equation}
Expanding this up to $O(1/m)$, we find
\begin{eqnarray}
   \Psi_{+v'}^{\rm LM}  &=&  \left[1 + i m \delta v \cdot x
    + \frac{\delta \slash v}{2} 
    + \frac{i}{4m} \Big( \delta \slash v ({\bigslash D} - v \cdot D)
        - D \cdot \delta v  \Big) + O \left(\frac{1}{m^2}\right) \right]
\Psi_{+v} \nonumber \\
        &=& \left[-{i \over 4m} D \cdot \delta v
        + O\left(1 \over m^2 \right)\right] \Psi_{+v'}^{\rm Ch}
\end{eqnarray}

\subsection{Velocity-operator notation}

Such a change in the reparameterization transformation may be induced in
a simple way, because besides transforming the fields, a
reparameterization transformation {\it also} changes the four-velocity $v^\mu$.

At this point, it is mnemonically useful to adopt a notation in which
the incorporation of different $v^\mu$ into the Hilbert space of the theory
is made explicit. This will make clear what happens when a
transformation that changes $v^\mu$ acts in the middle of a string of
operators that depend on $v^\mu$.

Define $\Psi_+$ to be a column vector consisting of all of the heavy
quark fields $\Psi_{+v}$. (The $+$ reminds us that the field is a HQET
field that satisfies $\slash v \Psi_+ = \Psi_+$. All four-velocities are
included in it, but not heavy antiquark fields, which would have to be
dealt with separately, though analogously).

Then $v^\mu$ may be treated as a four-vector {\it operator}
$\hat{v}^\mu$ that acts on $\Psi_+$. Its eigenspaces consist of states
of definite four-velocity with eigenvalue $v^\mu$. $\delta
\hat{v}^\mu(\epsilon_i)$ is also an operator. It commutes with
$\hat{v}^\mu$, and is defined in terms of the $\hat{v}^\mu$ operator via
the formula for the change in four-velocity under an infinitesimal
Lorentz transformation specified by the six infinitesimal parameters
$\epsilon_i$. (These could be boost rapidities and Euler angles, or
any other convenient parameterization. What
matters is that, unlike $\delta v^\mu$, they do not depend on the value
of $v^\mu$). It is the boost parameters which actually specify the
reparameterization transformation.

The shift in velocity is now accomplished explicitly by a shifting operator
$\hat{S}(\epsilon_i) = \delta_{v',v+\delta v(\epsilon_i)}$, which obeys the
commutation relations
\begin{eqnarray}
	[\hat{v}^\mu, \hat{S}(\epsilon_i)] &=&
	\hat{S} (\epsilon_i) \delta \hat{v}^\mu (\epsilon_i) \nonumber \\
	\left[\delta {\hat{v}}^\mu (\epsilon_i) , {\hat{S}}(\epsilon_i)\right]&=& 0
\end{eqnarray}
for infinitesimal $\epsilon_i$.
Now everything about a reparameterization transformation, including the shift
in four-velocity, is included in the action of the transformation operator
on the
field $\Psi_+$. We can handle both field and velocity transformations by
manipulating
operators in the usual way. 

\subsection{A field redefinition}

Rewritten in velocity-operator notation (with the hats on the various operators
omitted), Chen's reparameterization transformation to order $1 \over m$ is
\begin{eqnarray}
	M^{\rm Ch}(\epsilon_i) \Psi_+
	&=& S(\epsilon_i) \left[ 1 + i m \delta v(\epsilon_i) \cdot x + 
   \frac{\delta \slash v(\epsilon_i)}{2} \right. \nonumber \\
   && \left. + \frac{\delta \slash v(\epsilon_i)}{4m}
   i ({\bigslash D} - v \cdot D) + O\left(1 \over m^2 \right) \right] \Psi_+
\end{eqnarray}
and Luke and Manohar's is
\begin{eqnarray}
	M^{\rm LM}(\epsilon_i) \Psi_+ &=&
	S(\epsilon_i) \left[ 1 + i m \delta v(\epsilon_i) \cdot x + 
   \frac{\delta \slash v(\epsilon_i)}{2} \right. \nonumber \\
   && \left. + \frac{\delta \slash v(\epsilon_i)}{4m}
   i ({\bigslash D} - v \cdot D) -{i \over 4m} D \cdot \delta v(\epsilon_i)
   + O\left(1 \over m^2 \right)\right] \Psi_+
\end{eqnarray}
where $S(\epsilon_i)$, $v^\mu$, and $\delta v^\mu(\epsilon_i)$ are now
understood
to be operators.

Consider the field-redefinition operator
\begin{equation}
	R \Psi_+ = \left[1 -{i \over 4m} D \cdot v \right] \Psi_+
	\label{eq:redef}
\end{equation}
The important thing about $R$ is that it is completely independent of
$\epsilon_i$, so it
may be applied to $\Psi_+$ even in situations that have nothing to do with
reparameterization
transformations. It is a valid means of redefining fields so as to obtain
one formulation
of HQET from another.

Then applying $R$ to Chen's transformation reveals that
\begin{eqnarray}
	R M^{\rm Ch}(\epsilon_i) \Psi_+ &=& R M^{\rm Ch}(\epsilon_i) R^{-1} R
\Psi_+ \nonumber \\
	&=& M^{\rm Ch}(\epsilon_i) R \Psi_+
	+ S(\epsilon_i) \left[-{i \over 4m} D \cdot \delta v(\epsilon_i)\right] R
\Psi_+ \nonumber \\
	&=& M^{\rm LM}(\epsilon_i) R \Psi_+ + O \left(1 \over m^2 \right)
\end{eqnarray}
Even though the difference between Chen's transformation and Luke and
Manohar's appears
to depend on $\delta v(\epsilon_i)$, the shift-independent field redefinition
$R$ turns Chen's transformation into Luke and Manohar's, to order $1/m$.
The redefined field $R\Psi$ transforms under Luke and Manohar's
reparameterization
transformation.

The field redefinition $R$ is not a symmetry of the Lagrangian.
However, since it is proportional to $D \cdot v$, the changes that it
induces in the Lagrangian are manifestly
class II operators. In fact, it is precisely the
field redefinition necessary to absorb the class II operator
$-{1 \over 2m} \left(D \cdot v\right)^2$ in the Lagrangian obtained
from tree-level matching, when the Lagrangian is written in terms of
the field $R \Psi_+$.

The field redefinition necessary to absorb order $1/m$ and order $1/m^2$
class II operators in the Lagrangian obtained from tree-level matching is
\begin{equation}
	R' \Psi_+ = \left[1 - {iD \cdot v \over 4m} + {3(iD \cdot v)^2 \over 32m^2}
	\right] \Psi_+
	\label{eq:secondorderredef}
\end{equation}
In general, a derivative term at order $1 / m^j$ might
affect the form of the reparameterization transformation at order $ 1/m^{j-1}$,
because of the order $m$ term in the reparameterization transformation. In
this case, however, this does not happen, and the extra term in $R'$ has no
effect on the reparameterization transformation to order $1/m$.

(In \cite{Mci96} the field redefinition shown is the inverse
of (\ref{eq:secondorderredef}), because of notational conventions. Here
we define the new Lagrangian to be the original expression written in terms
of the transformed fields.)

Luke and Manohar's transformation, at least when expanded to first order in
$1/m$,
is a symmetry, not of the tree-level matching Lagrangian, but of the tree-level
Lagrangian with the class II operators removed. Chen's transformation, on
the other hand, is a symmetry of the tree-level Lagrangian with class II
operators included. In Appendix B, it is demonstrated that Chen's transformation
is not unique in this regard. There are other reparameterization transformations
that preserve the entire tree-level matching Lagrangian to all orders in $1/m$.

%=======================================================================
\section{Reparameterization invariance constraints on the effective
         Lagrangian}
%=======================================================================
Reparameterization invariance leads to important constraints for the
coupling constants in the effective Lagrangian. Due to the
possiblity of field redefinitions, neither the form of the Lagrangian
nor the form of the reparameterization transformation $\Psi_{+v} \to
\Psi_{+v'}$ is unique. However, a field redefinition such as $R$ above
will induce only class II terms in the Lagrangian, and cannot change
the constraints on the coefficients of class I terms in the Lagrangian.

\subsection{Chen's transformation}

Requiring invariance of the effective Lagrangian in (5) under
Chen's tranformation law leads to the following constraints
on the coefficients of the class I operators \cite{Mci96,Bal96,CKO95}
\begin{eqnarray} \label{eq:constr}
   C_{kin} & = & 1
\nonumber \\
   2 C_{mag} & = & C_2 + 1
\end{eqnarray}
In addition, it sets the following
constraint on the coefficients of some of the
class II operators:
\begin{equation} \label{eq:classIIconstr}
	C_a = C_{D \cdot v} + {1 \over 2}
\end{equation}
Note that in \cite{CKO95}, Chen, Kuang, and Oakes use an inconsisent
basis that is non-Hermitian and does not
include all of the necessary class II operators. 
They derive a spurious
reparameterization constraint equating a class I coefficient with a class II
coefficient ($c_4 = c_6$ in their paper). That this constraint is not gauge
invariant was noted in \cite{Bal96}.

\subsection{Luke and Manohar's transformation}

Luke and Manohar derived the same constraint for $C_{kin}$. When
discussing the relationship between $O_{mag}$ and $O_2$, they noted that
the combination
\begin{equation}
	O_{mag} + 2 O_2 + O\left(1 \over m^3 \right)
	\label{eq:combo}
\end{equation}
is reparameterization invariant, and that $O_{mag}$ is not related to
the leading-order Lagrangian by reparameterization invariance.

The second of these statements needs qualification.
$C_{mag}$ may be varied independently of the leading-order Lagrangian.
However, the presence of the leading-order Lagrangian does modify the
relationship between $C_{mag}$ and $C_2$, because the reparameterization
transformation acting on the leading-order Lagrangian yields a term at
$O(1/m)$
\begin{equation}
\delta {\cal L}_0 =
-{g \over 4m} {\overline{\Psi}_{+v}} \delta v_\mu \sigma^{\mu \nu} G_{\nu
\rho} v^\rho {\Psi_{+v}}
\end{equation}
which may only be cancelled by including a {\it difference} between
$C_{mag}$ and
$2 C_2$. This is why the constraint resulting from either Chen's
transformation or Luke and Manohar's is actually $2 C_{mag} = C_2 + 1$.
The reparameterization invariance of (\ref{eq:combo}) gives us the freedom
to change $C_{mag}$ and $C_2$ subject to this constraint without
violating reparameterization invariance.
It is easy to jump from the statements in \cite{Luk92} to the incorrect
conclusion that $C_2 = 2 C_{mag}$, but a close reading of \cite{Luk92}
reveals that Luke and Manohar never actually state this, and it is not
actually implied by what they do state. 
(Indeed, in \cite{Mci96}, two of us jumped to exactly that erroneous 
conclusion, and then incorrectly reasoned that results from
tree-level matching and one-loop running did not
agree with the class I constraints from Luke and Manohar's transformation).

Luke and Manohar's transformation also induces the class II constraint
\begin{equation}
	C_a = C_{D \cdot v}
\end{equation}
This does not agree with the result of the usual tree-level matching
procedure. However, it may be made to hold by a field redefinition, such
as the one which sets $C_a = C_{D \cdot v} = 0$.

\subsection{Field redefinitions}

A general field redefinition $\Psi'_v = R \Psi_{+v}$ which preserves
the projection property $P_v \Psi_{+v} = \Psi_{+v}$ and which transforms the
class I part of the general effective Lagrangian into itself has the form
\begin{equation}
   R \Psi_{+v} = 
   \left[1 + \frac{a}{2m} i v \cdot D +
   \frac{b}{4m^2} D^2 + \frac{c}{4m^2} \sigma_{\mu\nu} D^\mu D^\nu
   + \frac{d}{4m^2} (i v \cdot D)^2 + O(\frac{1}{m^3}) \right] \Psi_{+v}
\end{equation}
where $a$, $b$, $c$, $d$ are complex numbers.  The field redefinitions
(\ref{eq:redef})
and (\ref{eq:secondorderredef}) are redefinitions of this type.
It is straightforward to check that this transformation applied to the
general effective 
Lagrangian in (\ref{eq:eff}) does not change the class I terms. Applying
such redefinitions to a reparameterization symmetry $M(\epsilon_i)$ yields
a family
of reparameterization transformations $RM(\epsilon_i)R^{-1}$ which preserve
the class
I constraints.

This does not generalize to higher orders; at $1/m^3$, the coefficients
of the class I terms may change under field redefinitions unless the
form of the field redefinitions is restricted further (however, the
transformation
(\ref{eq:redef}) induces only class II terms to all orders).

%=======================================================================
\section{Loops, matching, and running}
%=======================================================================

The effective theory does not have the same short distance behavior as full QCD.
This must be taken into account by introducing suitable matching corrections. In
this section, we show that Chen's RPI symmetry still holds when this matching is
performed to order $1 \over m^2$, but to all orders in $\alpha_s$.

\subsection{Comparing with explicit running calculations}

It has been checked explicitly that renormalization of the effective
Lagrangian at one loop does fulfill the class I constraints in
(\ref{eq:constr}) \cite{Mci96,Bal96}. Other class I running calculations
(\cite{Blo96}, and the revised version of \cite{lee}) give equivalent
results for running of the class I operators at order $1/m^2$.
\cite{Bal94} does not report results for the running for $C_1$,
but does give a result for $C_2$ which agrees with the
above and with (\ref{eq:constr}).

In fact, the results for class II operators calculated in \cite{Bal96}
and \cite{Blo96}) also obey the class II constraint from Chen's transformation,
(\ref{eq:classIIconstr}). Translated into the operator basis of \cite{Bal96},
(\ref{eq:classIIconstr}) becomes \begin{equation} {1 \over 2} C_1^{(2)}
+ C_3^{(2)} - C_3^{(1)} = {1 \over 2} \end{equation} where we have used
$C_3^{(2)} = C_4^{(2)}$ by Hermiticity. This identity is satisfied at
tree level ($C_1^{(2)} = -1, C_3^{(2)} = 0, C_3^{(1)} = -1$ in their
operator basis). It is also maintained by their calculated one-loop
running, independently of their background field gauge-fixing parameter.
Agreement with (\ref{eq:classIIconstr}) is more manifest in \cite{Blo96},
since their class I operator basis is more similar to the one we are using.

This raises the question of whether these constraints are preserved more
generally under quantum corrections, beyond tree-level matching and
one-loop running.

\subsection{Spinors and 1PI Green's functions}

The general prescription for matching one theory to another at some
momentum scale is to ensure that the 1PI Green's functions of the two
theories describe the same physics at that scale, in an expansion in
inverse powers of the effective theory cutoff. The same transitions must
possess the same amplitudes when expanded in this way.

Spinors that appear on external legs of Feynman diagrams are always
solutions in momentum space of the unperturbed equation of motion. The
free field equations for the quark fields are different in QCD and HQET,
since parts of the quark-quark Green's function that arise from the
leading equation of motion in full QCD are attributed to higher-order
``interaction'' terms in HQET.

Therefore, the spinors one puts on external legs in QCD are not the same
as the ones used in HQET for the same physical situation. To find the Dirac
spinor in terms of the corresponding HQET spinor, one
substitutes $p^\mu = m v^\mu + k^\mu$ into the solutions of the momentum-space
free-field Dirac equation, and writes the resulting expression in terms of
a HQET spinor $u_{+v}$ for which $\slash v u_{+v} = u_{+v}$. For a HQET
spinor $u_{+v}$, the corresponding QCD spinor is
\begin{equation}
        u_{QCD} = \left[ 1 + {1 \over 2m + k \cdot v}
        \left(\slash k- k \cdot v \right) \right] u_{+v}
        \label{eq:spinormatch}
\end{equation}

The calculation may be simplified by putting external quark momenta on shell.
This is necessary so that, later, we can use form-factor decompositions to say
things about operator coefficients. It also transforms factors such as $k \cdot
v$ into higher-order quantities in $1 \over m^2$, simplifying the series
expansion to finite order. Taking $(m v + k)^2 = m^2$ for external quarks gives
the modified matching relation
\begin{equation}
        u_{QCD} = \left[ 1 + {1 \over 2m - k^2/(2 m)}
        \left(\slash k + {k^2 \over 2m} \right) \right] u_{+v}
        \label{eq:onshellmatch}
\end{equation}
and allows factors elsewhere in the 1PI Green's functions to be similarly
moved to higher orders in $1/m$.

This procedure has the disadvantage of slightly complicating the calculation of
coefficients of ``class II operators'' which vanish according to the free-field
equation of motion in the effective theory. It does not make it impossible
to say
anything about such operators, since we are making the spinors obey the full
theory's free-field equation of motion, rather than making the fields obey the
effective theory's coupled equation of motion. Some remnants of these operators
will remain, but putting the quark momenta on shell will give some Feynman
vertices of different class II operators the same form, so that we can only say
things about linear combinations of them.

\subsection{Gauge invariance}

When matching at tree level, it was possible to maintain gauge invariance
explicitly at all steps of the calculation. This is because, at tree level, the
generating functional of 1PI Green's functions is identical to the Lagrangian.
Therefore, one can match Lagrangians, deal with fields instead of spinors, and
use covariant derivatives instead of momenta.

When calculating loop diagrams, on the other hand, it is necessary to choose a
gauge. Gauge invariance can be made somewhat explicit by using background field
gauge, but diagrams will still treat interactions with different numbers of
gluons as separate vertices, and in the spinor-matching procedure above we treat
momenta separately from gluon couplings. The consequences of gauge invariance
then reappear later in the form of Ward identities relating different Green's
functions to one another. We will make use of one such identity when
proving that
Chen's RPI symmetry holds under loop matching corrections to order $1\over m^2$.

\subsection{Regularization scheme}

Since a matching prescription does not involve the infrared divergent
terms in a theory, the regularization scheme used for infrared
divergences does not matter, as long as it is used consistently in the
two theories. 
Thus we can use dimensional regularization
to regularize both ultraviolet and infrared divergences \cite{hpet4,neubert}. 
When used with
$\overline{\hbox{MS}}$, this eliminates all loop diagrams
that do not possess a mass scale other
than the renormalization scale $\mu$. This includes all loop diagrams in
HQET, since there the quark mass becomes a factor in coupling constants
rather than a contribution to the propagator.

Therefore, using this regularization scheme eliminates the need to
calculate HQET loop diagrams when matching to any order in perturbation
theory. We calculate 1PI loop diagrams to any desired order in full QCD,
with external quarks on shell and all divergences dimensionally
regularized; apply the spinor substitution (\ref{eq:onshellmatch}); and
adjust the couplings in the HQET Lagrangian so that the derived 1PI
Green's function arises {\it at tree level}.

Using this regularization scheme affords us an opportunity to prove
relations to all orders in $\alpha_s$. Lorentz and parity invariance of
full QCD allow us to write its 1PI Green's functions, with all loop
corrections included, in terms of invariant form factors. If the Green's
functions in HQET may be computed at tree level, then the structure of
the full QCD Green's functions directly implies constraints upon the
coupling constants of the HQET Lagrangian. To order $1 \over m^2$, it is
sufficient to consider the 1PI quark-quark and quark-quark-gluon Green's
functions in QCD.

\subsection{The quark two-point function}

The matching of the quark two-point function just corresponds to what we
already know about tree-level matching of free fields. Loops can only
yield mass and field renormalizations in the full theory, so after these
divergences have been subtracted off with counterterms, the amputated
1PI Green's function is
\begin{equation}
        i {\bar u}_{QCD} (\slash q - m) u_{QCD}
\end{equation}
where $q^\mu$ is the full momentum of the quark. Making the
substitution (\ref{eq:onshellmatch}), and simplifying the result
using the projection identity $\slash v u_{+v} = u_{+v}$,
yields the two-point function for HQET:
\begin{eqnarray}
    \Gamma_{{\bar q} q} &=& i \bar u_{+v} \left[ 1 + {1 \over 2m - k^2/(2 m)}
        \left(\slash k+ {k^2 \over 2m} \right) \right]
        (m \slash v+ \slash k- m) \nonumber \\
        &&\left[ 1 + {1 \over 2m - k^2/(2 m)}
        \left(\slash k+ {k^2 \over 2m} \right) \right] u_{+v}
\end{eqnarray}
This determines the coupling of every operator in HQET which
contains a two-quark Feynman vertex with no gluons. To order $1/m^2$,
applying the usual projection identities for heavy quark spinors,
it is simply
\begin{equation}
        i \bar u_{+v} \left(k \cdot v + {k^2 \over 2m} \right) u_{+v} +
O\left(1 \over m^3\right)
\end{equation}
and it ends up enforcing the RPI constraint $C_{kin} = 1$. It will also
constrain the coefficients of many high-order operators such as
${1 \over m^{2n-1}}{\overline{\Psi}_{+v}} (D^2)^n {\Psi_{+v}}$.

\subsection{The quark-quark-gluon three-point function}

The quark-quark-gluon vertex function is more interesting, because
there can be quantum corrections to the structure in $p^2$, where $p^\mu$ is
the transferred momentum. However, considerations of Lorentz invariance and
parity
limit the 1PI vertex function in a manner familiar from QED. There is a
Dirac form factor $F_1$ and a Pauli form factor $F_2$, which can depend on the
momenta only via $p^2$:
\begin{equation}
        \Gamma_{{\bar q} g q}^{\mu a}= ig \, {\bar u'} \gamma^\mu T^a u \,
F_1 (p^2,g,m,\mu)
        - {g \over 2m} \, {\bar u'} \sigma^{\mu \nu} T^a u \, p_\nu \, F_2
(p^2,g,m,\mu)
\end{equation}
Furthermore, $F_1 (p^2 = 0) = 1$, because of gauge invariance.
$F_2 (p^2 = 0)$, giving the ``anomalous chromomagnetic moment,'' is not
constrained by symmetry and can be affected by loop corrections.

Regularizing all loop divergences with dimensional regularization, making
the substitution (\ref{eq:onshellmatch}), and applying $\slash v u_{+v} =
u_{+v}$
as above gives the
{\it tree level} vertex function in HQET. To order $1/m^2$, where $p^\mu$ is the
transferred momentum and $k'^\mu$ is the final residual
momentum of the heavy quark, it is
\begin{eqnarray}
        &&i g \, {\bar u'}_{+v} T^a u_{+v} \, v^\mu \, F_1
          + {i g \over 2 m} \, {\bar u'}_{+v} T^a u_{+v} \, (2 {k'}^\mu -
p^\mu) \, F_1 \nonumber \\
        &&- {g \over 2 m} \, {\bar u'}_{+v} \sigma^{\mu \alpha} T^a u_{+v}
\, p_\alpha \, (F_1 + F_2) \nonumber \\
        &&+ {i g \over 8 m^2} \, {\bar u'}_{+v} T^a u_{+v} \, v^\mu p^2
\,(F_1 + 2 F_2)
          + {i g \over 8 m^2} \, {\bar u'}_{+v} T^a u_{+v} \, v^\mu [k'^2 +
(k'-p)^2] \, F_1 \nonumber \\
        &&+ {g \over 4 m^2} \, {\bar u'}_{+v} \sigma^{\alpha \beta} T^a
u_{+v} \,
             k'_\alpha p_\beta v^\mu \, (F_1 + 2 F_2)
\end{eqnarray}

The term that goes like $k^2 + (k-p)^2$ at order $1/m^2$ is a contribution
from class II operators. It looks like the Feynman vertices of $O_a$,
but because external momenta are on shell, it can also arise from $O_{D
\cdot v}$.
The remaining terms all come from the class I operators.

Expanding the
form factors as $F_i(p^2) = F_{i0} + {p^2/m^2} F_{i2} + O(1/m^4)$ makes it
possible to read off the coefficients directly, with some ambiguity in the
case of the class II operators:
\begin{eqnarray}
        C_{kin} &=& F_{10} \nonumber \\
        C_{mag} &=& F_{10} + F_{20} \nonumber \\
        C_1 &=& F_{10} + 8 F_{12} + 2 F_{20} \nonumber \\
        C_2 &=& F_{10} + 2 F_{20} \nonumber \\
		C_a - C_{D \cdot v} &=& {1 \over 2} F_{10} \nonumber \\
\end{eqnarray}
(Comparing with the Feynman rules listed in the long e-print version of
\cite{Mci96}, it is evident that terms in the Feynman vertices of $O_1$
and $O_2$ with factors of $p \cdot v$ have vanished here. 
This is again because
of the on-shell quark momenta, which promote these terms to order
$1/m^3$.)

This procedure yields no constraints on $C_1$, and, to this order, Chen's
RPI does not constrain it either. Applying the Ward identity $F_{10} = 1$
yields Chen's RPI constraints
\begin{eqnarray}
        C_{kin} &=& 1 \nonumber \\
        2 C_{mag} &=& C_2 + 1 \nonumber \\
        C_a &=& C_{D \cdot v} + {1 \over 2}
        \label{eq:chenconstraints}
\end{eqnarray}
Of course, the first relation already followed from the two-point function.
That it shows up here as well is a consequence of gauge symmetry.

\subsection{Running under the renormalization group}

In heavy quark effective field theory, we typically want to know the
values of coefficients at some momentum scale which is far below the
scale where matching to the full theory is done. After matching to the
full theory to some order in the number of loops, one uses the
renormalization group equation to determine how the coefficients in the
effective field theory Lagrangian evolve under large changes in scale.
The anomalous dimensions to use are typically calculated using diagrams
with one more loop than was used in matching.

As described in Section 6.1, it is known that running at one loop
preserves the reparameterization invariance constraints to order
$1/m^2$. The result derived above implies that the class I constraints should
apply for arbitrary numbers of loops. This is because renormalization
group running can be seen as a special case of the matching procedure,
which includes the terms from arbitrarily large orders in loops which
dominate when the scale is far below the matching scale. If the RPI
constraints apply at all orders in loops upon matching, they must
therefore also apply to the coefficients found by running under the
renormalization group. Therefore, we have shown not only that
class I reparameterization invariance constraints apply to order $1/m^2$
upon matching to full QCD, but that they apply under renormalization
group running as well, to all orders in $\alpha_s$.

Which transformation is actually a symmetry of the Lagrangian depends on
the class II terms, and therefore on how the quark fields are defined.
It is useful, as in \cite{Mci96}, to eliminate class II terms at all
stages of matching and running. One starts with the Lagrangian with
class II terms absorbed by a field redefinition. Then the
renormalization group running incorporates a field redefinition that
continuously absorbs class II terms induced by the running. Under these
conditions (if the class I operators are defined according to our
operator definitions), Luke and Manohar's transformation is a symmetry
of the Lagrangian to order $1/m^2$, since it is a symmetry of a
Lagrangian that satisfies the class I constraints and has no class II
terms.

On the other hand, if the heavy quark fields are defined by tree-level
matching in the usual way, and are not redefined to remove class II
operators (as in most existing renormalization calculations,
such as \cite{lee,Bal96,Blo96,Bal94}), then our results imply
that Chen's constraints on the class II coefficients also
hold under running, to order $1/m^2$ and to all orders in $\alpha_s$.
Then Chen's transformation is a symmetry of the renormalized Lagrangian
to order $1/m^2$.

%========================================================================
\section{Conclusions}
%========================================================================
The form of a reparameterization transformation may be modified by conjugating
it with other symmetry transformations, or with field redefinitions that
affect the coefficients of class II operators. We have demonstrated that the
forms of reparameterization invariance advocated by Chen and by Luke and
Manohar are both members of a the resulting family of viable reparameterization
transformations. Of the two, only Chen's is a member of the more restricted
family of symmetries of the entire Lagrangian derived from tree-level matching.

Both transformations induce the {\it same} constraints on class I operator\
coefficients to order $1/m^2$. We have proven that these constraints hold not
only at tree level, but to all orders in $\alpha_s$, upon matching between HQET
and QCD and under renormalization-group running. The transformations, in this
sense, are symmetries of the quantum theory as well as the classical theory. The
constraint imposed by Chen's transformation on the class II operators also holds
to order $1/m^2$ and all orders in $\alpha_s$, if the fields are not
redefined to
remove or modify the class II terms.

%========================================================================
\section{Acknowledgements}
%========================================================================
We would like to thank Markus A. Luty for very important discussions, and
for his notes on reparameterization invariance and
baryons in the $1/N$ expansion (unpublished). We would also like
to thank Thorsten Ohl and Christopher Balzereit for suggesting that we
consider the case of the class II constraints. Further thanks are due
to Pilar Hernandez, Aneesh V. Manohar, and Raman Sundrum.

M.F. would like to thank the members of the Theoretical
Physics Group at Harvard University for their kind hospitality.
This work is supported in part by the National Science Foundation
(Grant \#PHY-9218167) and by the Deutsche Forschungsgemeinschaft.

\appendix
%========================================================================
\section{Invariance of ${\cal L}^{\rm tree}$}
%========================================================================
The tree level matching Lagrangian is given by
\begin{equation}
   {\cal L }^{\rm tree} = \overline {\Psi}_{+v} A(v) \Psi_{+v}
\end{equation}
where
\begin{equation}
   A(v) = i v D + i \slash{D}_\perp \frac{1}{2 m + i v D} i \slash{D}_\perp
\end{equation} 
We want to prove invariance under the transformation
\begin{eqnarray}
   & & v  \to  v + \delta v
\nonumber \\
\nonumber \\ & &
   \Psi_{+v}  \to  \left[ 1 + i m \delta v x +
   \frac{\delta \slash{v}}{2} +
   \frac{\delta \slash{v}}{2} \frac{1}{2 m + i v D} i \slash{D}_\perp
   \right] \Psi_{+v}
\nonumber \\
\nonumber \\ & &
   \overline{\Psi}_{+v}  \to 
   \overline{\Psi}_{+v} \left[ 1 - im \delta v x +
   \frac{\delta \slash{v}}{2} +
   i \slash{D}_\perp \frac{1}{2 m + i v D}
   \frac{\delta \slash{v}}{2} \right]
\end{eqnarray}
Now
\begin{equation}
   \delta {\cal L} = \delta \overline{\Psi}_{+v} A \Psi_{+v}
   + \overline{\Psi}_{+v} A \delta \Psi_{+v} 
   + \overline{\Psi}_{+v} \delta A \Psi_{+v}
\end{equation}
Here
$$
   P_{+v} \delta A P_{+v} 
   = P_{+v} \delta v^\mu \Big[ i D_\mu - i D_\mu \frac{1}{2 m 
   + i v D} i \slash{D}_\perp - i \slash{D}_\perp \frac{1}{2 m
   + i v D} i D_\mu \frac{1}{2 m + i v D} i \slash{D}_\perp
$$
$$
\qquad
- i \slash{D}_\perp \frac{1}{2 m + i v D} i D_\mu \Big] P_{+v}
$$
\begin{equation}
  = 
   P_{+v} \left\{
   i \delta v D 
   - i \slash{D}_\perp \frac{1}{2 m
   + i v D} i \delta v D \frac{1}{2 m + i v D} i \slash{D}_\perp
   \right\} P_{+v}
\end{equation}
where we have used
\begin{equation}
   \frac{\partial}{\partial v^\mu} \frac{1}{2m + i v D}
  = \frac{1}{2m + i v D} (- i D^\mu) \frac{1}{2m + i v D}
\end{equation}
Now consider
\begin{equation}
   \delta \overline{\Psi}_{+v} A \Psi_{+v} +
   \overline{\Psi}_{+v} A \delta \Psi_{+v} 
   = \overline{\Psi}_{+v} \{ [A,i m \delta v x] + A \delta M +
   \overline{ \delta M} A \} \Psi_{+v}
\end{equation}
where
\begin{eqnarray}
   \delta M & = & \frac{\delta \slash{v}}{2}
   + \frac{\delta \slash{v}}{2} \frac{1}{2 m + i v D} i \slash{D}_\perp
\end{eqnarray}
Now
\begin{equation}
   [A, i m \delta v x] = [i v D 
   + i \slash{D}_\perp \frac{1}{2 m + i v D} i \slash{D}_\perp,
   i m \delta v x]
\end{equation}
Firstly 
\begin{eqnarray}
   [i v D, i m \delta v x] = [i v \partial, i m \delta v x] 
   - g [v A, i m \delta v x] = 0
\end{eqnarray}
because of $v \delta v = 0$. Secondly we have
\begin{eqnarray}
   [i \slash{D}_\perp, i m \delta v x] 
   = [ i \slash{D}, i m \delta v x] 
\nonumber \\ 
\nonumber \\ 
   = [ i \slash{\partial}, i m \delta v x]
   = - m \delta \slash{v}
\end{eqnarray}
Thirdly
\begin{equation}
  [ \frac{1}{2 m + i v D}, i m \delta v x] = 0
\end{equation}
And so 
we obtain
$$
[A, i m \delta v x] = 
[ i \slash{D}_\perp \frac{1}{2 m + i m D} i \slash{D}_\perp, 
   i m \delta v x]
$$
\begin{equation} \qquad
  = - m i \slash{D}_\perp \frac{1}{2 m + i v D} \delta \slash{v}
    - m \delta \slash{v} \frac{1}{2 m + i m v D} i \slash{D}_\perp
\end{equation}
Furthermore
$$
   A \delta M =
   i v D \frac{\delta \slash{v}}{2} + \frac{\delta \slash{v}}{2}
   \frac{i v D}{2 m + i v D} i \slash{D}_\perp
   + i \slash{D}_\perp \frac{1}{2 m + i v D} i \slash{D}_\perp
      \frac{\delta \slash{v}}{2}
$$ 
\begin{equation} \qquad
   + i \slash{D}_\perp \frac{1}{2 m + ivD}i \slash{D}_\perp
   \frac{\delta \slash{v}}{2} \frac{1}{2 m + i v D} i \slash{D}_\perp
\end{equation}
Now 
$$
  P_{+v} i v D \frac{\delta \slash{v}}{2}P_{+v} = 0
$$
$$
   P_{+v} 
 i \slash{D}_\perp \frac{1}{2 m + i v D} i \slash{D}_\perp
      \frac{\delta \slash{v}}{2} P_{+v}
%$$
%$$
= 
   P_{+v} 
 i \slash{D}_\perp \frac{1}{2 m + i v D} i P_{-v} \slash{D}_\perp P_{-v}
      \frac{\delta \slash{v}}{2} P_{+v}
$$
\begin{equation}
= 
  P_{+v} i \slash{D}_\perp \frac{-ivD}{2m+ivD} \delta \slash{v} P_{+v}
\end{equation}
and so
$$
   P_{+v} A \delta M P_{+v} =
  P_{+v} \Big[ \frac{\delta \slash{v}}{2} \frac{ivD}{2m+ivD} 
                 i \slash{D}_\perp
   - i \slash{D}_\perp \frac{ivD}{2m+ivD} \delta \slash{v}
$$
\begin{equation} \qquad
   + i \slash{D}_\perp \frac{1}{2m+ivD} i \slash{D}_\perp
     \frac{\delta \slash{v}}{2} \frac{1}{2m+ivD} i \slash{D}_\perp
   \Big] P_{+v}
\end{equation}
Similarly
$$
   P_{+v} \overline{\delta M} A P_{+v} =
   P_{+v} \Big[ - \delta \slash{v} \frac{ivD}{2m+ivD} i \slash{D}_\perp
   + i \slash{D}_\perp \frac{ivD}{2m+ivD} \frac{\delta \slash{v}}{2}
$$
\begin{equation} \qquad
  + i \slash{D}_\perp \frac{1}{2m+ivD} \frac{\delta \slash{v}}{2}
    i \slash{D}_\perp \frac{1}{2m+ivD} i \slash{D}_\perp
\end{equation}
and finally (sandwiching between a pair of $P_{+v}$'s is implied)
\begin{eqnarray}
  A \delta M + \overline{\delta M} A & = &
   - \frac{\delta \slash{v}}{2} \frac{ivD}{2m+ivD} i \slash{D}_\perp
   - i \slash{D}_\perp \frac{ivD}{2m+ivD}\frac{\delta\slash{v}}{2}
\nonumber \\
\nonumber \\
& & 
   + i \slash{D}_\perp \frac{1}{2m+ivD} i \delta v D \frac{1}{2m+ivD} 
     i \slash{D}_\perp
\end{eqnarray}
Plugging everything together, we find the variation of the Lagrangian:
\begin{eqnarray}
   \delta {\cal L} & = & \overline{\Psi}_{+v} \Big\{
   - m i \slash{D}_\perp \frac{1}{2m+ivD} \delta \slash{v}
   - m \delta \slash{v} \frac{1}{2m+ivD} i \slash{D}_\perp
   - \frac{\delta \slash{v}}{2} \frac{ivD}{2m+ivD} i \slash{D}_\perp
\nonumber \\
\nonumber \\ & &
   - i \slash{D}_\perp \frac{ivD}{2m+ivD}\frac{\delta \slash{v}}{2}
   + i \slash{D}_\perp \frac{1}{2m+ivD} i \delta v D \frac{1}{2m+ivD}
     i \slash{D}_\perp
   + i \delta v D 
\nonumber \\
\nonumber \\ & &
   - i \slash{D}_\perp \frac{1}{2m+ivD} i \delta v D
     \frac{1}{2m+ivD} i \slash{D}_\perp
   \Big\} \Psi_{+v}
\nonumber \\
\nonumber \\ & = &
  \overline{\Psi}_{+v} \Big\{
    i \delta v D - \frac{\delta \slash{v}}{2} \frac{ivD + 2m}{2m+ivD}
      i \slash{D}_\perp
   - i \slash{D}_\perp \frac{ivD+2m}{2m+ivD}\frac{\delta \slash{v}}{2}
  \Big\} \Psi_{+v}
\nonumber \\
\nonumber \\ & = &
  \overline{\Psi}_{+v} \Big\{
   i \delta v D - i \delta v D
  \Big\} \Psi_{+v} = 0
\end{eqnarray}
which concludes the proof.
%---------------------------------------------------------
\section{Is Chen's transformation unique?}
%---------------------------------------------------------
In this appendix we  will show that Chen's transformation law is not
unique
even in the restricted family of 
symmetries of the full Lagrangian derived from 
tree-level matching.

So let's try to find the class of all reparameterization transformation which
leave the Lagrangian 
\begin{equation}
   {\cal L}^{\rm tree}(v,\Psi_{+v}) = \overline{\Psi}_{+v} A(v) \Psi_{+v}
\end{equation}
invariant ($A(v)$ has been given in the previous section). So consider an
infinitesimal transformation
\begin{equation}
   v \to v' = v + \delta v
\end{equation}
with
\begin{equation}
   \delta v \cdot v = 0
\end{equation}
What is the most general ansatz for the transformation $\Psi_{+v} \to
\Psi_{+v'}$?
The field $\Psi_{+v'}$ must have two properties: (i) it must have the correct
projection property $P_{+v'} \Psi_{+v'} = \Psi_{+v'}$ and (ii) the derivative 
acting on it must produce the correct residual momentum $k'$ instead of $k$.
The most general ansatz compatible with these two requirements is
\begin{equation}
   \Psi_{+v'} = \Big[ 1 + i m \delta v \cdot x \Big]
   P_{+v'} B \Psi_{+v}
\end{equation}
where $B$ must not depend explicitly on $x$, but is otherwise arbitrary. $B$
is a matrix in Dirac space and will contain covariant derivatives.
Define
\begin{equation}
   B_\pm := P_{\pm v} B
\end{equation}
Then
\begin{equation}
   \Psi_{+v'} = \left[ \left(1 + i m \delta v \cdot x + 
   \frac{\delta \slash v}{2} \right) B_+ + \frac{\delta \slash v}{2} B_-
   \right] \Psi_{+v}
\end{equation}
For $\delta v \to 0$, we must have $\Psi_{+v'} = \Psi_{v}$.
Therefore we can assume $B_-$ to be of lowest order in $\delta v$, i.e.
$B_- = O(\delta v)^0$ and $B_+ = 1 + \delta B_+$, where $\delta B_+ = 
O(\delta v)$. And so
\begin{equation}
  \Psi_{+v'} = \left[ 1 + \delta B_+ + i m \delta v \cdot x +
   \frac{\delta \slash v}{2} + \frac{\delta \slash v}{2} B_- \right] 
   \Psi_{+v}
\end{equation}
At this point one can notice that the only combination of $B_+$ and
$B_-$ which enters the transformation law is
$$
   \delta B_+ + \frac{\delta \slash v}{2} B_-
$$
but not $B_+$ or $B_-$ themselves. 
%Therefore I am not sure if this seperation
%into $B_\pm$ is really necessary / helpful. See below!

Using the various tricks and techniques of the previous section, one can
now inserte this general transformation into the Lagrangian, and require
its variation to vanish. Defining 
\begin{equation}
   C_- := B_- - \frac{1}{2m+iv\cdot D} i {\bigslash D}_\perp 
\end{equation}
this finally leads to
\begin{equation}
   0 = \delta {\cal L} =
   \overline{\Psi}_{+v} \left\{ A(v) \left[ \delta B_+ + \frac{\delta \slash v}
   {2} C_- \right] + \left[\overline{\delta B_+} + \overline{C_-} 
   \frac{\delta \slash v}{2} \right] A(v) \right\} \Psi_{+v}
\end{equation}
A solution to this equation is $\delta B_+ = 0$ and $C_- =0 $. This is Chen's
transformation. Are there other solutions?

Firstly note again, that only $\delta B_+ + \frac{\delta{v}}{2} B_-$
enters in the transformation law, i.e. only the sum 
 $\delta B_+ + \frac{\delta{v}}{2} C_-$ matters and solutions with
 $\delta B_+ + \frac{\delta{v}}{2} C_- = 0$ do not lead to different
reparameterization transformations.

Secondly, however, there are solutions with
 $\delta B_+ + \frac{\delta{v}}{2} C_- \neq 0$. A non-trivial example
is 
\begin{equation}
   \delta B_+ + \frac{\delta \slash v}{2} C_- = i \frac{i \delta v \cdot D}
   {m} \frac{A(v)}{m}
\end{equation}
Note that this transformation is 'class II' in a generalized sense, i.e.
it vanishes for classical solutions of the {\em full} tree level effective
Lagrangian with $A(v) \Psi_{+v} = 0$.

%=======================================================================

%=======================================================================

\end{document}